\documentstyle[aps, prl, twocolumn]{revtex}

\def\Z{{\bf Z}}
\def\F{{\bf F}}

\def\qed{$\Box$}

\begin{document}

\preprint{LAUR98-5842}

\title{How to share a quantum secret} 

\author{Richard Cleve$^1$\thanks{Email: {\tt cleve@cpsc.ucalgary.ca}},
Daniel Gottesman$^2$\thanks{Email: {\tt gottesma@t6-serv.lanl.gov}},
Hoi-Kwong Lo$^3$\thanks{Email: {\tt hkl@hplb.hpl.hp.com}}}

\address{$^1$ Department of Computer Science, University of Calgary,
Calgary, Alberta, Canada T2N 1N4\\
$^2$ T-6 Group, Los Alamos National Lab, Los Alamos, NM 87545\\
$^3$ Hewlett-Packard Labs, Bristol, UK BS34 8QZ}

\maketitle

\begin{abstract}
We investigate the concept of quantum secret sharing.  In a $((k,n))$ 
threshold scheme, a secret quantum state is divided into $n$ shares such 
that any $k$ of those shares can be used to reconstruct the secret, but 
any set of $k-1$ or fewer shares contains absolutely no information about 
the secret.  We show that the only constraint on the existence of
threshold schemes comes from the quantum ``no-cloning theorem'', which
requires that $n < 2k$, and, in all such cases, we give an efficient
construction of a $((k,n))$ threshold scheme.  We also explore
similarities and differences between quantum secret sharing schemes
and quantum error-correcting codes.  One remarkable difference is
that, while most existing quantum codes encode pure states as pure
states, quantum secret sharing schemes must use mixed states in some
cases.  For example, if $k \le n < 2k-1$ then any $((k,n))$ threshold
scheme {\em must\/} distribute information that is globally in a mixed
state.
\end{abstract}


Suppose that the president of a bank wants to give access to a 
vault to three vice presidents who are not entirely trusted.
Instead of giving the combination to any one individual, it may be 
desirable to distribute information in such a way that no vice president 
alone has any knowledge of the combination, but any two of them can 
jointly determine the combination.
In 1979, Blakely~\cite{Blakely} and Shamir~\cite{Shamir} addressed a
generalization of this problem, by showing how to construct schemes
that divide a secret into $n$ shares such that any $k$ of those shares 
can be used to reconstruct the secret, but any set of $k-1$ or fewer 
shares contains absolutely no information about the secret.  This is
called a {\em $(k,n)$ threshold scheme}, and is a useful tool for
designing cryptographic key management systems.

Now, consider a generalization of such schemes to the setting of 
{\em quantum} information, where the secret is an arbitrary unknown 
quantum state.
Salvail~\cite{Salvail} (see also~\cite{HBB}) obtained 
a method to divide an unknown qubit into two shares, each of which 
individually contains no information about the qubit, but which
jointly can be used to reconstruct the qubit.
Hillery, Bu\v zek, and Berthiaume~\cite{HBB} proposed a method for 
implementing some {\em classical\/} threshold schemes that uses 
quantum information to transmit the shares securely in the presence 
of eavesdroppers.

Define a {\em $((k,n))$ threshold scheme}, with $k \leq n$,
as a method to encode and divide an arbitrary {\em secret\/} quantum 
state (which is given but not, in general, explicitly known) into $n$ 
{\em shares} with the following two properties. First, from any $k$ 
or more shares the secret quantum state can be perfectly reconstructed.
Second, from any $k-1$ or fewer shares, no information {\em at all\/} 
can be deduced about the secret quantum state.
%%FOR QPH
Formally, this means that the reduced density matrix of these $k-1$ 
shares (with the other shares traced out) is independent of the value 
of the secret.
%%END QPH
Each share can consist of any number of qubits (or higher-dimensional 
states), and not all shares need to be of the same size. 
%%FOR QPH
In this paper 
we do not consider the problem of securely creating and distributing 
the shared secret, and simply assume that it can be done when necessary.
%%END QPH

Quantum secret sharing schemes might be used in the context of sharing
quantum keys, such as those proposed by Weisner~\cite{Wiesner} for
uncounterfeitable ``quantum money.''  They can also be used to provide
interesting ways of distributing quantum entanglement and nonlocality.
For example, suppose that Alice has one qubit of an EPR pair and a 
$((2,2))$ threshold scheme is applied to the other qubit to produce
a share for Bob and a share for Carol.  Then Alice and Bob together
have a product state (i.e., $\rho_{AB} = \rho_A \otimes \rho_B$), 
as do Alice and Carol; however, Bob and Carol can jointly construct a 
qubit from their shares that is in an EPR state with Alice's qubit.
Also, for quantum storage or quantum computations to
be robust in the worst-case situation where a component or a group of
components fail (due to sabotage by malicious parties or due to
defects), quantum secret sharing may prove to be a useful
concept. Finally, by definition, quantum secret sharing distributes
trust between various parties and prevents a small coalition of
malicious parties from learning a quantum secret.

Let us begin with an example of a $((2,3))$ threshold scheme.
The secret here is an 
arbitrary three-dimensional quantum state (a quantum trit or {\em qutrit}).
The encoding maps the secret qutrit to three qutrits as 
\begin{eqnarray}
\label{3trit}
\alpha\left| 0 \right\rangle + \beta\left| 1 \right\rangle + 
\gamma\left| 2 \right\rangle & \mapsto &
\alpha (\left|000 \right\rangle + \left|111 \right\rangle + 
\left|222 \right\rangle) + \\
& & \beta  (\left|012 \right\rangle + \left|120 \right\rangle + 
\left|201 \right\rangle) + \nonumber \\
& & \gamma (\left|021 \right\rangle + \left|102 \right\rangle + 
\left|210 \right\rangle), \nonumber
\end{eqnarray}
and each resulting qutrit is taken as a share.  Note that, from a
single share, absolutely no information can be deduced about the
secret, since each individual share is always in the totally mixed
state (an equal mixture of $\left|0 \right\rangle$, $\left|1
\right\rangle$, and $\left|2 \right\rangle$).  On the other hand, the
secret can be reconstructed from any two of the three shares as
follows.  If we are given the first two shares (for instance), add the
value of the first share to the second (modulo three), and then add
the value of the second share to the first, to obtain the state
\begin{equation}
(\alpha\left|0 \right\rangle + \beta\left|1 \right\rangle + 
\gamma\left|2 \right\rangle)
(\left|00 \right\rangle + \left|12 \right\rangle + \left|21 \right\rangle).
\end{equation}
The first qutrit now contains the secret.
%%FOR QPH
The reconstruction procedure for the other cases is similar, 
by the symmetry of mapping (\ref{3trit}) with respect to cyclic 
permutations of the three qutrits.
%%END QPH

%%FOR QPH
Note that, because the data is quantum, one must be careful
not to individually measure the shares while performing 
the reconstruction, since this will collapse any superposition 
of the basis states.
The same considerations arise with quantum 
error-correcting codes~\cite{shor9,steane7}.
In fact, the above example is a three-qutrit quantum code that can 
correct one erasure error.
%%END QPH
%%FOR PRL
%Note that the above example is similar to a quantum error-correcting 
%code~\cite{shor9,steane7}. In fact, it is a three-qutrit quantum code 
%that can correct one erasure error.
%%END PRL
Every quantum secret sharing scheme is, in some sense, a quantum 
error-correcting code; however, some error-correcting codes are 
not secret sharing schemes, since they may contain sets of shares 
from which {\em partial\/} information 
about the secret can be obtained.
For example, consider a four-qubit code \cite{prevent,Grassl} that 
corrects one erasure by the encoding
%%FOR QPH
\[
\alpha\left|0 \right\rangle + \beta\left|1 \right\rangle \longmapsto 
\alpha(\left|0000 \right\rangle + \left|1111 \right\rangle) + 
\beta(\left|0011 \right\rangle + \left|1100 \right\rangle)
\]
(the code can actually be extended to encode two qubits, but we do not
need this for our illustration).
%%END QPH
%%FOR PRL
%\[
%\alpha\left|0 \right\rangle + \beta\left|1 \right\rangle \longmapsto 
%\alpha(\left|0000 \right\rangle + \left|1111 \right\rangle) + 
%\beta(\left|0011 \right\rangle + \left|1100 \right\rangle).
%\]
%%END PRL
While it is true that any three
qubits suffice to reconstruct the secret, it is {\em not\/} true that
two qubits provide no information.  For instance, given the first and
third qubits, one can distinguish between the secrets $\left|0
\right\rangle$ and $\left|1 \right\rangle$.
%%FOR QPH
More generally, from
these two qubits, statistical information about the relative values of
$|\alpha|$ and $|\beta|$ can be obtained.
%%END QPH
Later, we shall show how to
obtain a $((3,4))$ threshold scheme with four qubits using a different
approach.

Returning to the $((2,3))$ threshold scheme using qutrits, note 
that it can be used to share a secret that is a qu{\em bit\/} by simply 
not using the third dimension of the input space (though the resulting 
shares are still full qutrits).
It turns out that there does not exist a $((2,3))$ threshold scheme 
for qubits in which each share is also a qubit.
This is because such a scheme would also be a three-qubit code 
that corrects single qubit erasure errors, which has been shown not 
to exist~\cite{Grassl}.

The $((2,3))$ qutrit threshold scheme can be used to construct a 
$((2,2))$ threshold scheme, by simply discarding (i.e., tracing out) 
one of the three shares.
Note that the resulting $((2,2))$ scheme produces a mixed state encoding 
even when the secret is a pure state.
%%FOR QPH
The encoding procedure can be defined by the following linear map 
on density matrices
\begin{eqnarray}
\left| 0 \right\rangle\!\left\langle 0 \right| & \mapsto & 
\left| 00 \right\rangle\!\left\langle 00 \right| + 
\left| 11 \right\rangle\!\left\langle 11 \right| + 
\left| 22 \right\rangle\!\left\langle 22 \right| \nonumber \\
\left| 1 \right\rangle\!\left\langle 1 \right| & \mapsto & 
\left| 01 \right\rangle\!\left\langle 01 \right| + 
\left| 12 \right\rangle\!\left\langle 12 \right| + 
\left| 20 \right\rangle\!\left\langle 20 \right| \\
\left| 2 \right\rangle\!\left\langle 2 \right| & \mapsto & 
\left| 02 \right\rangle\!\left\langle 02 \right| + 
\left| 10 \right\rangle\!\left\langle 10 \right| + 
\left| 21 \right\rangle\!\left\langle 21 \right|. \nonumber 
\end{eqnarray}
%%END QPH
Call a scheme that encodes pure state secrets using global pure states 
a {\em pure state scheme}, and a scheme for which the encodings 
of pure states are sometimes in global mixed states a {\em mixed state 
scheme}.
We shall show later that there does not exist a pure state $((2,2))$ 
threshold scheme.

On the other hand, if we do not insist on
protecting an arbitrary
secret, we could use the encoding
\begin{equation}
\alpha \left|0 \right\rangle + \beta \left|1 \right\rangle \mapsto 
\alpha (\left|00 \right\rangle - \left|11 \right\rangle) + 
\beta (\left|01 \right\rangle + \left|10 \right\rangle) .
\end{equation}
For the restricted set of secrets where $\alpha \cdot \beta^*$ is
real-valued, it functions as a $((2,2))$ threshold scheme.  However,
without this restriction, this is not a secret sharing scheme, since
(for example) it can be verified that a single share can completely
distinguish between the secrets $\left|0 \right\rangle + i\left|1
\right\rangle$ and $\left|0 \right\rangle - i\left|1 \right\rangle$.
Although such a scheme may be useful in some contexts, we shall
henceforth consider only ``unrestricted'' secret sharing schemes.

Note that the previously mentioned technique of discarding a share from 
a $((2,3))$ threshold scheme to obtain a $((2,2))$ threshold scheme 
(suggested by \cite{Lane} in the context of a different scheme) 
generalizes considerably:

\smallskip \noindent
{\bf Theorem 1.}
{\sl {From} any $((k,n))$ threshold scheme with $n > k$, a $((k,n-1))$ 
threshold scheme can be constructed by discarding one share.}
\smallskip

In the classical case, a $(k,n)$ threshold scheme exists for 
every value of $n \ge k$. However, this does not hold in the quantum
case, due to the 
%%FOR QPH
quantum ``no-cloning theorem''~\cite{WZ,Dieks},
which states that no operation can produce
multiple copies of an unknown arbitrary quantum state.
%%END QPH
%%FOR PRL
%``no-cloning theorem''~\cite{WZ,Dieks}.
%%END PRL

\smallskip \noindent
{\bf Theorem 2.}
{\sl If $n \ge 2k$ then no $((k,n))$ threshold scheme exists.}

\smallskip \noindent
{\bf Proof.}
If a $((k,n))$ threshold scheme exists with $n \ge 2k$ then the 
following procedure can be used to make two independent copies 
of an arbitrary quantum state (that is, to clone).
First, apply the $((k,n))$ scheme to the state to produce $n$ shares.
Then, taking two disjoint sets of $k$ shares, reconstruct two independent 
copies of the state.
This contradicts the ``no-cloning theorem'' \cite{WZ,Dieks}.
\hfill \qed
\smallskip

The five-qubit quantum code proposed in~\cite{BDSW,fivequbit} 
immediately yields a $((3,5))$ threshold scheme.
First, since it corrects any two erasure errors, it enables the secret 
to be reconstructed from any three shares.
Also, any pair of qubits provides no information about the data.
This is a consequence of the following more general theorem.
 
\smallskip \noindent 
{\bf Theorem 3.}  
{\sl If a quantum code with codewords of length $2k-1$ corrects $k-1$
erasure errors (which, for stabilizer codes~\cite{stab,CRSS}, is a
{$[[2k-1, 1, k]]_q$} code, where $q$ is the dimensionality of each
coordinate and of the~encoded state) then it is also a $((k,2k-1))$
threshold scheme.}

\smallskip \noindent
{\bf Proof.}
First, suppose that we are given a set of $k$ shares.
Since this set excludes precisely $k-1$ shares and the code 
corrects any $k-1$ erasures, the secret can be reconstructed from 
these $k$ shares.
On the other hand, suppose that we are given a set of $k-1$ shares.
This subset excludes a set of $k$ shares, from which we know 
that the secret can be perfectly reconstructed.
Now, in quantum mechanics, it is well-known that any information gain
on an unknown quantum state necessarily leads to its
disturbance~\cite{disturbance}.
Therefore, if a measurement on the given $k-1$ shares provided any 
information about the secret, then this measurement would disturb 
the information that the remaining $k$ qubits contain about the 
secret. This leads to a contradiction.
\hfill \qed \smallskip

Combining Theorem~3 with Theorem~1, we obtain

\smallskip \noindent
{\bf Corollary 4.}
{\sl {From} a $[[2k-1, 1, k]]_q$ code, a $((k,n))$ threshold scheme 
can be constructed for any $n < 2k$.}
\smallskip

For example, from the aforementioned five-qubit code, a $((3,4))$ 
threshold scheme and  $((3,3))$ threshold scheme can be obtained 
(by discarding shares).

Next, we prove the converse of Theorem~2.

\smallskip \noindent
{\bf Theorem 5.}
{\sl If $n < 2k$, then a $((k,n))$ threshold scheme exists.
Moreover, the dimension of each share can be bounded above by 
$2\max (2k-1,s)$, where $s$ is the dimension of the quantum secret.}

\smallskip \noindent
{\bf Proof.}
The proof is based on a class of {\em quantum polynomial codes}, which 
are similar to those defined by Aharonov and Ben-Or~\cite{AB}, who used 
them in the context of fault-tolerant quantum computation.
We will show how to construct such a code of length $m$ and degree $k-1$ 
whenever $m < 2k$, and that the data that it encodes can always 
be recovered from any $k$ of its $m$ coordinates. Then, considering the 
special case where $m=2k-1$, we obtain a $[[2k-1, 1, k]]_q$ code, for 
which Corollary~4 applies to prove the theorem.

Let $k$ and $m$ be given with $m < 2k$, and let $s$ be the dimension of the 
quantum state to be encoded.  Choose a prime $q$ such that $\max (m, s) 
\le q \le 2\max (m, s)$ (which is always possible~\cite{prime}) and let 
$\F = \Z_q$. For $c = (c_0,c_1,\ldots,c_{k-1}) \in \F^k$, define the 
polynomial $p_c(t) = c_0 + c_1 t + \cdots + c_{k-1} t^{k-1}$.  Let 
$x_0,\ldots,x_{m-1}$ be $m$ distinct elements of $\F$.  Encode a $q$-ary 
quantum state by the linear mapping which is defined on basis states 
$\left|s\right\rangle$ (for $s \in \F$) as
\begin{equation}
\label{encode}
\left|s \right\rangle \mapsto 
\sum_{{c \in \F^k} \atop {c_{k-1} = s}} 
\left| p_c(x_0),\ldots,p_c(x_{m-1}) \right\rangle.
\end{equation}
As an example, it turns out that mapping~(\ref{3trit}) (for the 
$((2,3))$ threshold 
%%FOR PRL
%scheme)
%%ENDPRL
%%FOR QPH
scheme given at the beginning of this paper)
%%END QPH
is a quantum polynomial code with $k=2$, $m = 3$, and $q=3$.

It now suffices to show that, given an encoding~(\ref{encode}) of a 
quantum state, the state can be recovered from any $k$ of the $m$ 
coordinates.
One way to show this is to apply the theory of CSS 
codes~\cite{CS,Steane}, noting that this code is formed from the 
two classical codes
\begin{eqnarray}
C_1 & = & \left\{ (p_c(x_0),\ldots,p_c(x_{m-1}))\,|\,c \in \F^k \right\} \\
C_2 & = & \left\{ (p_c(x_0),\ldots,p_c(x_{m-1}))\,|\,c \in \F^k, c_{k-1} = 0 
\right\}
\end{eqnarray}
and that ${\rm min}({\rm dist}\ C_1, {\rm dist}\ C_2^\perp) = m-k+1$.
{From} this it follows that the code corrects $m-k$ erasure errors.

For completeness, we also give an explicit decoding procedure for 
the case of interest, where $m=2k-1$.
We begin with some preliminary definitions.
For an invertible $d \times d$ matrix $M$, define the operation 
{\em apply $M$} to a sequence of $d$ quantum registers as applying 
the mapping
%%FOR QPH
\begin{equation}
\left|(y_0,\ldots,y_{d-1}) \right\rangle \mapsto 
\left|(y_0,\ldots,y_{d-1})M \right\rangle
\end{equation}
(where we are equating $\left|(y_0,\ldots,y_{d-1}) \right\rangle$ with 
$\left|y_0,\ldots,y_{d-1} \right\rangle$).
%%END QPH
%%FOR PRL
%\begin{equation}
%\left|(y_0,\ldots,y_{d-1}) \right\rangle \mapsto 
%\left|(y_0,\ldots,y_{d-1})M \right\rangle.
%\end{equation}
%%END PRL
For $z_0,\ldots,z_{d-1} \in \F$, define the $d \times d$ 
Vandermonde matrix
\begin{equation}
\left[V_d(z_0, \ldots, z_{d-1})\right]_{ij} = z_j^i
\end{equation}
(for $i, j \in \{0,\ldots,d-1\}$).  
%%FOR QPH
This matrix is invertible whenever
$z_0,\ldots,z_{d-1}$ are distinct.
%%END QPH
Also, note that applying
$V_d(z_0,\ldots,z_{d-1})$ to registers in state
$\left|c_0,\ldots,c_{d-1} \right\rangle$ yields the state
$\left|p_c(z_0),\ldots,p_c(z_{d-1}) \right\rangle$, where $c =
(c_0,\ldots,c_{d-1})$.

The secret can be recovered from any $k$ coordinates by the following 
procedure.
Call the $m$ registers containing the coordinates $R_0,\ldots,R_{m-1}$, 
and suppose that we are given, say, the first $k$ registers (that is, 
$R_0,\ldots,R_{k-1}$).
\begin{enumerate}
\item
Apply $V_k(x_0,\ldots,x_{k-1})^{-1}$ to $R_0,\ldots,R_{k-1}$.
\item
Cyclically shift the first $k$ registers by one to the right by 
setting $(R_0,R_1\ldots,R_{k-1})$ to $(R_{k-1},R_0,\ldots,R_{k-2})$.
\item
Apply $V_{k-1}(x_k,\ldots,x_{m-1})$ to $R_1,\ldots,R_{k-1}$.
\item
For all $i \in \{1,\ldots,k\!-\!1\}$, add $R_0 \!\cdot\! (x_{k+i-1})^{k-1}$ 
to $R_i$.
\end{enumerate}

Consider an execution of the above procedure on a state
resulting from the encoding~(\ref{encode}) on a basis state
$| s \rangle$.
After steps 1 and 2, the state of the $n$ registers is
\begin{eqnarray}
\label{after12}
\lefteqn{\sum_{c \in \F^k \atop c_{k-1} = s} 
\left|c_{k-1},c_0,\ldots,c_{k-2} \right\rangle
\left|p_c(x_k),\ldots,p_c(x_{m-1}) \right\rangle} 
& & \nonumber \\
& = &
\left|s \right\rangle\sum_{c \in \F^k \atop c_{k-1} = s} 
\left|c_0,\ldots,c_{k-2} \right\rangle
\left|p_c(x_k),\ldots,p_c(x_{m-1}) \right\rangle.
\end{eqnarray}
If the data is a basis state $\left|s \right\rangle$ (for some $s \in
\F$) then, at this point, its recovery is complete.  However, for a
general secret, which is a superposition of $\left|s \right\rangle$
states, register $R_0$ is entangled with the other registers.  The
entanglement is due to the fact that, in (\ref{after12}), the value of
$s$ can be determined by the value of any of the kets
$\left|c_0,\ldots,c_{k-2} \right\rangle
\left|p_c(x_k),\ldots,p_c(x_{m-1}) \right\rangle$.
%%FOR QPH
In fact, if we had
$m \ge 2k$ then $s$ could be determined from just the state of the
last $m-k$ registers, so it would be impossible to perform the
necessary disentanglement by accessing only the first $k$ registers.
Since $m = 2k-1$, this is not a problem and the remaining steps
correctly extract the data in the following manner.
%%END QPH
%%FOR PRL
%The remaining steps complete the extraction of the data.
%%END PRL

After steps 3 and 4, the state is 
\begin{eqnarray}
\label{after4}
\lefteqn{\left|s \right\rangle\sum_{c \in \F^k \atop c_{k-1} = s} 
\left|p_c(x_k),\ldots,p_c(x_{m-1}) \right\rangle
\left|p_c(x_k),\ldots,p_c(x_{m-1}) \right\rangle} \hspace*{5mm} 
& & \nonumber \\
&  =  & \left|s \right\rangle\sum_{y \in \F^{k-1}}  
\left|y_1,\ldots,y_{k-1} \right\rangle
\left|y_{1},\ldots,y_{k-1} \right\rangle, \hspace*{10mm}
\end{eqnarray}
where the last equality holds since, for any $s \in \F$ and 
$y_1,\ldots,y_{k-1} \in \F$, there is a unique $c \in \F^k$ 
with $c_{k-1} = s$ such that $p_c(x_{k+i-1}) = y_i$, for all 
$i \in \{1,\ldots,k-1\}$.
%%FOR QPH
Since the state of $R_1,\ldots,R_{m-1}$ is now independent of $s$, 
the decoding procedure is now correct for arbitrary data.
%%END QPH
%%FOR PRL
%The decoding procedure is now correct for arbitrary data.
%%END PRL
\hfill \qed \smallskip

Although we have focused on threshold schemes, 
it is possible to consider more general access structures.  In a
general quantum secret sharing scheme, from certain {\em authorized
sets} of shares, the secret can be reconstructed, while, from all other
sets of shares, no information can be obtained about the secret.
%%FOR QPH
Those other sets are called {\em unauthorized sets}.
%%END QPH
For example, consider a scenario with three shares, $A$, $B$, $C$, where
the authorized sets are $\{A,B\}$, $\{A,C\}$, and any superset of one
of these sets.  Such a secret sharing scheme can be easily implemented
by starting with the $((3,4))$ threshold scheme and bundling the first
two shares into the share $A$.

We have already seen relationships between quantum secret sharing
schemes and quantum error-correcting codes.
We now explore this connection more deeply.

The following proposition follows naturally from the usual formulation
of the conditions for a quantum error-correcting code.

\smallskip \noindent
{\bf Proposition 6.}
{\sl Let ${\cal C}$ be a subspace of a Hilbert space ${\cal H}$.  The
following conditions are equivalent:
\begin{enumerate}

\item[a)] ${\cal C}$ corrects erasures on a set $K$ of coordinates.

\item[b)] For any orthonormal basis $\{|\phi_i\rangle\}$ of ${\cal C}$,
\begin{eqnarray}
\langle \phi_i | E | \phi_j \rangle & = & 0\ \ \ (i \ne j)
\label{condb1} \\
\langle \phi_i | E | \phi_i \rangle & = & c(E) \label{condb2}
\end{eqnarray}
for all operators $E$ acting on $K$.

\item[c)] For all (normalized) $|\phi\rangle \in {\cal C}$ and all $E$
acting on $K$,
\begin{equation}
\langle \phi | E | \phi \rangle = c(E).
\label{condc}
\end{equation}

\end{enumerate}
}

Note that the same function $c(E)$ appears in conditions (b) and (c),
and that it is independent of $|\phi\rangle$ or $|\phi_i\rangle$.

\smallskip \noindent
{\bf Proof.}
a) $\Leftrightarrow$ b) is essentially the standard quantum error correction
conditions~\cite{BDSW,KL} applied to erasure errors~\cite{Grassl}.
b) $\Leftrightarrow$ c) is straightforward.
Alternately, a) $\Leftrightarrow$ c) follows from the main theorem 
of~\cite{NC}.
\hfill \qed \smallskip

Equation~(\ref{condb1}) says that in correcting errors, we will never
confuse two different basis vectors.  Equation~(\ref{condb2}) says
that learning about the error will never give us any information about
which basis vector we have.
%%FOR QPH
This is important, since that information
would constitute a measurement, collapsing a superposition of basis
vectors.
%%END QPH

On the other hand, condition~(\ref{condc}) simply says that the
environment can never gain any information about the state.  In other
words, the proposition tells us that protecting a state from noise is
exactly the same as preventing the environment from learning about it. 

Condition~(\ref{condc}) is also very convenient for our purposes,
since the two constraints that arise on a quantum secret sharing
scheme are the ability to correct erasures and the requirement that no
information be gained by unauthorized sets of shares.

%%FOR QPH
In the theory of quantum error-correcting codes, we usually consider
shares of the same dimension.  In contrast, in quantum secret sharing,
we would like to allow shares to live in Hilbert spaces of different
sizes. Nevertheless, it is still true that conditions a), b), and c)
in Proposition~6 are equivalent.
%%END QPH

\smallskip \noindent
{\bf Theorem 7.}
{\sl An encoding $f: |\psi\rangle \mapsto |\phi\rangle$ is a pure state
quantum secret sharing scheme iff
%%FOR PRL
%equation~(\ref{condc}) holds
%%END PRL
%%FOR QPH
\begin{equation}
\langle \phi | E | \phi \rangle = c(E)
\label{condition}
\end{equation}
%%END QPH
(independent of $|\phi\rangle$) whenever $E$ is an operator acting
on the complement of an authorized set or when $E$ is an
operator acting on an unauthorized set.}
\smallskip

%%FOR QPH
For instance, for the three-qutrit scheme~(\ref{3trit}) and
$E_j\left|y_1,y_2,y_3 \right\rangle = \omega^{y_j}\left|y_1,y_2,y_3
\right\rangle$, where $\omega = \exp (2 \pi i/3)$, we have $\langle
\phi | E_j | \phi \rangle = 0$ for all states $|\phi\rangle$ used in
the scheme.
%%END QPH

\smallskip \noindent
{\bf Proof.}
Let ${\cal C}$ be the image of $f$.
$S$ is an authorized set iff the subspace ${\cal C}$ can correct for
erasures on $K$, the complement of $S$.  By Proposition~6, this means
$S$ is an authorized set iff 
%%FOR QPH
(\ref{condition}) 
%%END QPH
%%FOR PRL
%(\ref{condc})
%%END PRL
holds for all $E$
acting on $K$.  $T$ is an unauthorized set whenever we can gain no
information about the state $|\psi\rangle$ from any measurement on
$T$.  That is, the expectation value $\langle \phi | E | \phi \rangle$
is independent of $|\phi\rangle \in {\cal C}$ for any operator $E$ we
could choose to measure, which means it must act on $T$.  Again, this
is 
%%FOR QPH
condition~(\ref{condition}).
%%END QPH
%%FOR PRL
%condition~(\ref{condc}).
%%END PRL
\hfill \qed \smallskip

Theorem~7 has at least one remarkable consequence:

\smallskip \noindent
{\bf Corollary 8.}
{\sl For a pure state quantum secret sharing scheme, every unauthorized 
set of shares is the complement of an authorized set and vice-versa.}

\smallskip \noindent
{\bf Proof.}
If the complement of an authorized set of shares $S_1$ were another 
authorized set $S_2$ then we could create two copies of the secret 
from $S_1$ and $S_2$, violating the no-cloning theorem.
Therefore, the complement of an authorized set is always an 
unauthorized set.

On the other hand, by Proposition~6, if
%%FOR QPH
condition~(\ref{condition})
%%END QPH
%%FOR PRL
%condition~(\ref{condc})
%%END PRL
holds on an unauthorized set $T$, we can correct erasures on $T$, and
therefore reconstruct the secret on the complement of $T$.  Therefore,
the complement of an unauthorized set is always an authorized set.
\hfill \qed \smallskip

For a pure state $((k,n))$ threshold scheme, this condition implies that 
$n-k = k-1$.
Therefore:

\smallskip \noindent
{\bf Corollary 9.}
{\sl Any $((k,n))$ pure state threshold scheme satisfies $n = 2k - 1$.}

\smallskip 

Clearly, this corollary does not apply to mixed state schemes, since 
we have constructed $((k,n))$ threshold schemes with $n < 2k-1$.
  
We would like to thank Dorit Aharonov, Alexei Ashikhmin, Charles
Bennett, Andr\'e Berthiaume, Vladimir Bu\v zek, H. F. Chau, Mark
Hillery, Brendan Lane, Debbie Leung, and Louis Salvail for helpful
discussions.  Part of this work was completed during the 1998
Elsag-Bailey -- I.S.I.  Foundation research meeting on quantum
computation, and the 1998 meeting at the Benasque Center for Physics.
R.C. is supported in part by Canada's NSERC.  D.G. is supported by the
Department of Energy under contract W-7405-ENG-36.


\begin{thebibliography}{99}
\bibitem{Blakely} G.~Blakely, ``Safeguarding cryptographic keys,''
Proc.\ AFIPS {\bf 48}, 313--317 (1979).
\bibitem{Shamir} A.~Shamir, ``How to share a secret,'' Communications
of the ACM, {\bf 22}, 612--613 (1979).
\bibitem{Salvail} L.~Salvail, private communication.
\bibitem{HBB} M.~Hillery, V.~Bu\v zek, and A.~Berthiaume, ``Quantum secret
sharing,'' quant-ph/9806063.
\bibitem{Wiesner} S.~Wiesner, ``Conjugate Coding,'' SIGACT News {\bf 15},
pp. 78--88 (1983).
\bibitem{shor9} P.~Shor, ``Scheme for reducing decoherence in quantum
memory,'' Phys.\ Rev.\ A {\bf 52}, 2493--2496 (1995).
\bibitem{steane7} A.~M.~Steane, ``Error correcting codes in quantum
theory,'' Phys.\ Rev.\ Lett.\ {\bf 77}, 793--797 (1996).
\bibitem{prevent} L.~Vaidman, L.~Goldenberg, and S.~Wiesner, ``Error
prevention scheme with four particles,'' Phys. Rev. A {\bf 54}, 1745--1748
(1996); quant-ph/9603031.
\bibitem{Grassl} M.~Grassl, T.~Beth, and T.~Pellizzari, ``Codes for
the quantum erasure channel,'' Phys.\ Rev.\ A {\bf 56}, 33--38 (1997);
quant-ph/9610042.
\bibitem{Lane} B.~Lane, personal communication (1997).
\bibitem{WZ} W.~K.~Wootters and W.~H.~Zurek, ``A single quantum cannot
be cloned,'' Nature {\bf 299}, 802--803 (1982).
\bibitem{Dieks} D.~Dieks, ``Communication by EPR devices,''
Phys.\ Lett.\ A {\bf 92}, 271--272 (1982).
\bibitem{BDSW} C.~Bennett, D.~DiVincenzo, J.~Smolin, and W.~Wootters,
``Mixed state entanglement and quantum error correction,'' Phys.\
Rev.\ A {\bf 54}, 3824--3851 (1996); quant-ph/9604024.
\bibitem{fivequbit} R.~Laflamme, C.~Miquel, J.~P.~Paz, and W.~Zurek,
``Perfect quantum error correction code,'' Phys.\ Rev.\ Lett.\ {\bf
77}, 198--201 (1996); quant-ph/9602019.
\bibitem{stab} D.~Gottesman, ``Class of quantum
error-correcting codes saturating the quantum Hamming bound,'' Phys.\
Rev.\ A {\bf 54}, 1862--1868 (1996); quant-ph/9604038.
\bibitem{CRSS} A.~R.~Calderbank, E.~M.~Rains, P.~W.~Shor, and
N.~J.~A.~Sloane, ``Quantum error correction and orthogonal geometry,''
Phys.\ Rev.\ Lett.\ {\bf 78}, 405--408 (1997); quant-ph/9605005.
\bibitem{disturbance} C. H. Bennett, G. Brassard and N. David Mermin, ``Quantum
Cryptography without Bell's Theorem,''
Phys.\ Rev.\ Lett.\ {\bf 68}, 557--559 (1992).
\bibitem{AB} D.~Aharonov and M.~Ben-Or, ``Fault-tolerant quantum
computation with constant error,'' {\it Proc.\ 29th Ann.\ ACM Symp.\ on 
Theory of Computing}, pp.~176--188 (ACM, New York, 1998); quant-ph/9611025.
\bibitem{prime} M.~Aigner, G.~M.~Ziegler, {\it Proofs from The Book}, 
pp.~7--12 (Springer, Berlin, 1998).

See also S.~Ramanujan, ``A proof of Bertrand's Postulate,'' 
{\it J.\ of the Indian Math.\ Soc.} {\bf 11}, pp.~181--182 (1919).
%P.~Erd\"os, Acta Sci.\ Math.\ (Szeged) {\bf 5} 194 (1930--32).
\bibitem{CS} A.~R.~Calderbank and P.~W.~Shor, ``Good quantum
error-correcting codes exist,'' Phys.\ Rev.\ A {\bf 54}, 1098--1105 
(1996); quant-ph/9512032.
\bibitem{Steane} A.~Steane, ``Multiple particle interference and
quantum error correction,'' Proc.\ Roy.\ Soc.\ Lond.\ A {\bf 452},
2551--2577 (1996); quant-ph/9601029.
\bibitem{NC} M. A. Nielsen and C. M. Caves, ``Reversible quantum
operations and their application to teleportation,'' Phys.\ Rev.\ A
 {\bf 55}, pp 2547--2556 (1997); quant-ph/9608001
\bibitem{KL} E.~Knill and R.~Laflamme, ``A theory of quantum
error-correcting codes,'' Phys.\ Rev.\ A {\bf 55}, 900--911 (1997);
quant-ph/9604034.
\end{thebibliography}
\end{document}